\documentclass[prl,twocolumn,superscriptaddress,floatfix,preprintnumbers,amssymb,amsmath,showpacs]{revtex4}
\usepackage{graphicx}
\usepackage{dcolumn}
\usepackage{bm}
\usepackage[latin1]{inputenc}
\usepackage[mathscr]{eucal}
\usepackage{epsfig}
\begin{document}
\title{Information entropic superconducting microcooler }
\author{A. O. Niskanen}
\affiliation{CREST-JST, Kawaguchi, Saitama 332-0012,Japan}
\affiliation{VTT Technical Research Centre of Finland, Sensors, PO BOX 1000, 02044 VTT, Finland}
\author{Y. Nakamura}
\affiliation{CREST-JST, Kawaguchi, Saitama 332-0012,Japan}
\affiliation{NEC Fundamental Research Laboratories, Tsukuba, Ibaraki 305-8501, Japan}
\affiliation{The Institute of Physical and Chemical Research (RIKEN), Wako, Saitama 351-0198, Japan}
\author{J. P. Pekola}
\affiliation{Low Temperature Laboratory, Helsinki University of Technology, PO BOX 3500, 02015 TKK, Finland}
\pacs{74.50.+r,85.80.Fi,03.67.-a}


\date{\today}
\begin{abstract}
We consider a design for a cyclic microrefrigerator using a
superconducting flux qubit. Adiabatic modulation of the flux
combined with thermalization can be used to transfer energy from a
lower temperature normal metal thin film resistor to another one at
higher temperature. The frequency selectivity of photonic heat
conduction is achieved by including the hot resistor as part of a
high frequency LC resonator and the cold one as part of a
low-frequency oscillator while keeping both circuits in the
underdamped regime. We discuss the performance of the device in an
experimentally realistic setting. This device illustrates the
complementarity of information and thermodynamic entropy as the
erasure of the quantum bit directly relates to the cooling of the
resistor.
\end{abstract}
\maketitle

For the purpose of quantum computing, the coherence properties of
superconducting quantum bits (qubits) should be optimized by decoupling them 
from all noise sources as well as possible. However, many
interesting experiments can be envisioned also when the decoupling
is far from perfect. One such experiment closely related to
coherence optimization is using a qubit as a spectrometer
\cite{astafiev,bertet,yoshihara} for the environmental noise by
monitoring the effect of the environment on the quantum two-level
system. Here we focus on the opposite phenomenon, i.e. the effect of
a qubit on the environment. Recently a superconducting flux qubit
\cite{mooij,chiorescu} with a quite small tunneling energy from the
point of view of quantum computing was cooled using sideband cooling
and a third level \cite{mit} from about 400 mK down to 3 mK.
Motivated by this experiment we consider the possibility of using a
single quantum bit as a cyclic refrigerator for environmental
degrees of freedom. The utilized heat conduction mechanism is
photonic which was recently studied also in experiment
\cite{meschke}. Besides the possible practical uses, the device is
interesting physically as it directly illustrates the connection
between information entropy and thermodynamical entropy. For related
superconducting high-frequency cooler concepts see eg.
Refs.~\cite{hauss,nisrfset}.
\begin{figure}
\begin{picture}(150,220)
\put(-50,-50){\includegraphics[width=0.50\textwidth]{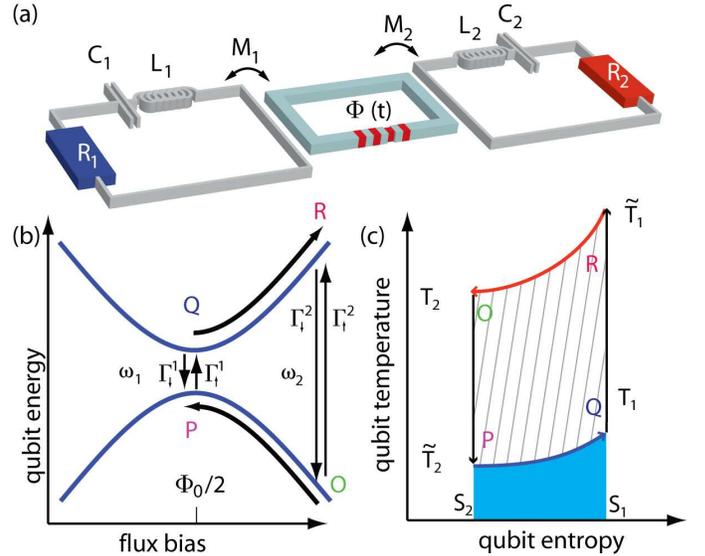}}
\end{picture}
\caption{\label{fg:scheme}
(color online) Principle of the flux-qubit cooler. 
(a) Layout of the circuit. (b) Energy band diagram.
(c) Schematic of the cooling cycle in the qubit temperature-entropy plane.
}
\end{figure}

Here we study a flux qubit coupled inductively to two
different loops shown in Fig.~\ref{fg:scheme}a. In loop $j$
($j=1,2$) we have a resistor $R_j$ in series with an inductance
$L_j$ and a capacitance $C_j$. These form two damped harmonic
oscillators. The resistors are in general at different temperatures
$T_1$ and $T_2$. The coupling of the qubit to these two admittances
$Y_1$ and $Y_2$ is assumed to be sufficiently large to dominate the
relaxation of the qubit. This assumption can be easily validated by
e.g. increasing the mutual inductance. The flux qubit is an
otherwise superconducting loop except for three or four Josephson
junctions with suitably picked parameters. In particular one of the
junctions is made smaller than others to form a two-level system.
When biased close to half of the flux quantum $\Phi_0=h/2e$, the
qubit can be described (in persistent current basis) by the
Hamiltonian
\begin{equation}\label{eq:ham}
H/\hbar=-\frac{1}{2}\left(\Delta\sigma_x+\varepsilon\sigma_z\right)
\end{equation}
where $\sigma_x$ and $\sigma_z$ are Pauli matrices,
$\hbar\varepsilon=2I_{{\rm p}}(\Phi-\Phi_0/2)$ is the flux-tunable energy bias and $\Phi$ is the controllable flux
threading the qubit loop. Away from $\Phi=\Phi_0/2$ the eigenstates have the persistent currents $\pm I_{{\rm p}}$
circulating in the loop. The tunneling energy $\hbar\Delta$ results in an anticrossing at $\Phi=\Phi_0/2$ and there
the energy eigenstates do not carry average current. The resonant angular frequency of the qubit is $\omega=\sqrt{\varepsilon^2+\Delta^2}$.

Consider the ideal cycle shown in Fig.~\ref{fg:scheme}b-c where the
bias of the flux qubit is swept slowly (slower than $\Delta/2\pi$)
between two extreme values $\varepsilon_1$ and $\varepsilon_2$
corresponding to two different energy level separations
$\hbar\omega_1$ and $\hbar\omega_2$. Let us further assume that
$\omega_j\approx \omega_{LCj}$ and $Q_j\gg 1$, where
$\omega_{LCj}=1/\sqrt{L_jC_j}$ and $Q_j=\sqrt{L_j/C_j}/R_j$. This
choice guarantees that the qubit mainly couples to resistor $R_1$
($R_2$) at bias point 1 (2). The cooling cycle consists of steps O,
P, Q and R. First in step O the qubit has the angular frequency
$\omega_2$ and is allowed to thermalize. Because of the bandwidth
limitations imposed by the reactive elements, the qubit tends to
thermalize with resistor $R_2$ to temperature $T_2$. In the next
step P the flux bias is adiabatically changed to point 1 such that
the level populations do not change but the energy eigenstates do.
The sweep is assumed to be however faster than relaxation. In point
1 the angular frequency is reduced to $\omega_1$. Because the level
populations and therefore the Boltzmann factors do not change the
qubit must now be at lower temperature $\tilde{T}_2$ given by
$\tilde{T}_2=T_2\omega_1/\omega_2$ in order to compensate for the
change of the qubit splitting. Note that the quantum mechanical
adiabaticity implies also thermodynamical adiabaticity: while the
energy eigenbasis changes the level populations and thus also
entropy do not change. In step Q the qubit is allowed to thermalize
to temperature $T_1$ which results in heating of the qubit and in
cooling of resistor 1 if $\tilde{T}_2<T_1$. At this point the
ideally pure quantum state of the qubit gets erased and information
stored is lost. The entropy of the qubit increases, but locally the
entropy of resistor 1 decreases such that one can say that some
information is ``stored'' in the resistor as it cools but naturally
with some loss. Finally in step R the qubit is adiabatically shifted
back to frequency $\omega_2$ which results in heating of the qubit
to the effective temperature $\tilde{T}_1=T_1\omega_2/\omega_1$
which is assumed to be higher than $T_2$. The excess energy is
dumped to admittance 2 when the cycle starts again from the
beginning. Note that due to the condition $\tilde{T}_2<T_1$ resistor
1 can never be cooled below $T_2\omega_1/\omega_2$. Since there is no isothermal stage in the above 
cycle the present device is not even in principle a Carnot cooler but rather an Otto-type device.\cite{quan}

The density matrix of the qubit with the resonant angular frequency $\omega$ at temperature $T$ ($\beta=(k_{\rm B}T)^{-1}$) is
given by
\begin{equation}
\rho_{\rm eq}(\beta,\varepsilon)=\frac{1}{2}\left[I+\left(\frac{\Delta}{\omega}\sigma_x+
\frac{\varepsilon}{\omega}\sigma_z\right)\tanh\left(\frac{\beta\hbar\omega}{2}\right)\right].
\end{equation}
Using this the cooling power and the efficiency of the ideal cycle
in Fig.~\ref{fg:scheme}c can be easily calculated. It is given by
the area of the shaded region in the entropy-temperature plane below
points P and Q. In principle one could solve for the effective
temperature of the qubit along the line between points P and Q as a
function of entropy given by $S=-k_{\rm B}{\rm Tr}(\rho\ln \rho)$.
Alternatively, we can simply note that the expectation value of the
energy stored in the qubit in point P is $E_P={\rm Tr}(\rho_{\rm
eq}(\beta_2\omega_2/\omega_1,\varepsilon_1)H_1)$ while after
relaxation we have $E_Q={\rm Tr}(\rho_{\rm
eq}(\beta_1,\varepsilon_1)H_1)$, where $H_1=H(\varepsilon_1)$ is the
Hamiltonian at point 1. We thus get for the ideal cooling power
\begin{equation}\label{eq:pow}
P/f=E_Q-E_P=\frac{\hbar\omega_1 e^{-\beta_1\hbar\omega_1}}{e^{-\beta_1\hbar\omega_1}+1}
-\frac{\hbar\omega_1 e^{-\beta_2\hbar\omega_2}}{e^{-\beta_2\hbar\omega_2}+1} \leq \frac{\hbar\omega_1}{2},
\end{equation}
where $f$ is the pump frequency. The cooling power achieves the
maximum value of $\hbar\omega_1f/2$ when the thermal population in
step O (and P) is small and when the population in step Q is large,
i.e. when $\beta_2 \hbar\omega_2 \gg 1$ and $\beta_1 \hbar\omega_1
\ll 1$. Naturally a practical device has to be designed to fulfill
the first condition always, in which case the smallest achievable
temperature is on the order of $\hbar\omega_1/k_{\rm B}$ below which
the cooling power decreases rapidly. The dynamic range could be made
wider by a tunable $\Delta$ which can be achieved by splitting the
smallest junction into a dc SQUID geometry. 
Another figure of merit is
the ratio $\eta$ of the heat removed
from resistor 1 divided by the heat added to resistor
2. It can be obtained as the ratio of the shaded area divided by
the sum of the hatched area and the shaded area,
i.e., $\eta=(E_Q-E_P)/(E_R-E_O)$ where
$E_O={\rm Tr}(\rho_{\rm eq}(\beta_2,\varepsilon_2)H_2)$ and
$E_R={\rm Tr}(\rho_{\rm eq}(\beta_1\omega_1/\omega_2,\varepsilon_2)H_2)$. This simplifies
neatly to $\eta=\omega_1/\omega_2<1$ which is in harmony with the second law of thermodynamics.

For more quantitative analysis we have to consider the details of
the relaxation rates due to the baths. The Golden Rule transition
rates due to resistor $j$ are given by
\begin{eqnarray}
\Gamma_{\downarrow,\uparrow}^j&=&\frac{2\pi}{\hbar^2}|\langle0|dH/d\Phi|1\rangle|^2 M_j^2S_I^j(\pm \omega_j) \nonumber \\
&=&\frac{2\pi}{\hbar^2}\frac{I_{\rm p}^2\Delta^2}{\omega^2}M_j^2S_I^j(\pm \omega_j)
\end{eqnarray}
where the positive sign corresponds to relaxation. The total
thermalization rate is $\Gamma_{\rm
th}^j=\Gamma_\uparrow^j+\Gamma_\downarrow^j$. Here the
unsymmetrized noise spectrum is given by
\begin{eqnarray}
  S_I^j(\omega)&=&\frac{1}{2\pi}\int_{-\infty}^\infty e^{-i\omega t}\langle\delta I_j(0)\delta I_j(t) \rangle dt \nonumber  \\
&=&\frac{1}{2\pi}\left(\frac{2\hbar\omega {\rm Re}Y_j(\omega)}{1-\exp(-\beta_j\hbar\omega)}\right).
\end{eqnarray}
where
${\rm Re}Y_j(\omega)=R_j^{-1}/[1+Q_j^{2}(\frac{\omega}{\omega_{LCj}}-\frac{\omega_{LCj}}{\omega})^2]$
is the real part of admittance of circuit $j$.
The total relaxation rate is thus
\begin{equation}
\Gamma_{\rm th}^j=\frac{2(I_{\rm p}\Delta M_j)^2\coth\left(\frac{\beta_j\hbar\omega}{2}\right)}
{R_j\hbar\omega\left(1+Q_j^{2}\left(\frac{\omega}{\omega_{LCj}}-\frac{\omega_{LCj}}{\omega}\right)^2\right)}.
\end{equation}
To model the behavior of the device we utilize the Bloch master equation (see e.g Ref.~\cite{makhlin})
given in our case by
\begin{equation}\label{eq:bloch}
\dot{\vec{M}}=-\vec{B}\times\vec{M}-\Gamma_{\rm
th}^1(\vec{M}_\parallel-\vec{M}_{T_1}) -\Gamma_{\rm
th}^2(\vec{M}_{\parallel}-\vec{M}_{T_2})-\Gamma_2\vec{M}_\perp,
\end{equation}
where $\vec{M}={\rm Tr}(\vec{\sigma}\rho)$ is the ``magnetization''
of the qubit, and $\vec{B}={\Delta}\vec{x}+\varepsilon\vec{z}$ is
the fictitious magnetic field. Note however that the z-component of
$\vec{B}$ and $\vec{M}$ do correspond to real magnetic field and
magnetization, respectively. In Eq.~(\ref{eq:bloch})
$\vec{M}_\parallel$ and $\vec{M}_\perp$ are the components of the
magnetization parallel and perpendicular to $\vec{B}$, respectively.
These are explicitly
\begin{eqnarray}
&\vec{M}_\parallel=\frac{1}{\omega^2}(\Delta M_x+\varepsilon M_z)({\Delta}\vec{x}+\varepsilon\vec{z}) \\
&\vec{M}_\perp=\frac{\varepsilon^2 M_x-\Delta\varepsilon M_z}{\omega^2}\vec{x}+M_y\vec{y}+
\frac{\Delta^2 M_z-\Delta\varepsilon M_x}{\omega^2}\vec{z}.
\end{eqnarray}
Here $\vec{M}_{T}$ stands for the $\varepsilon$-dependent
equilibrium magnetization of a qubit at temperature $T$ given
explicitly by
\begin{eqnarray}
\vec{M}_{T}=\left(\frac{\Delta}{\omega}\vec{x}+
\frac{\varepsilon}{\omega}\vec{z}\right)\tanh\left(\frac{\beta\hbar\omega}{2}\right)
\end{eqnarray}
and ${\Gamma_2=(\Gamma_{\rm th}^1+\Gamma_{\rm
th}^2)/2+\Gamma_\varphi}$ is the dephasing rate. The possibility of
pure dephasing at the rate $\Gamma_\varphi$ has been included. In
the simulation we neglect pure dephasing due to the intentionally
large dominating thermalization rate. Equation~(\ref{eq:bloch})
describes relaxation towards instantaneous equilibrium with two
competing rates due to two different thermal baths. Equations of
this type are usually used in the stationary case, but for driving
frequencies slower than $\Delta/\hbar$ it should be also valid. As
is obvious from Eq.~(\ref{eq:bloch}), the qubit actually tends to
relax towards an effective $\varepsilon$-dependent equilibrium
magnetization ${(\Gamma_{\rm th}^1\vec{M}_{T_1}+\Gamma_{\rm
th}^2\vec{M}_{T_2})/(\Gamma_{\rm th}^1+\Gamma_{\rm th}^2)}$ at the
rate $\Gamma_{\rm th}^1+\Gamma_{\rm th}^2$.

\begin{figure}
\begin{picture}(150,200)
\put(-60,0){\includegraphics[width=0.55\textwidth]{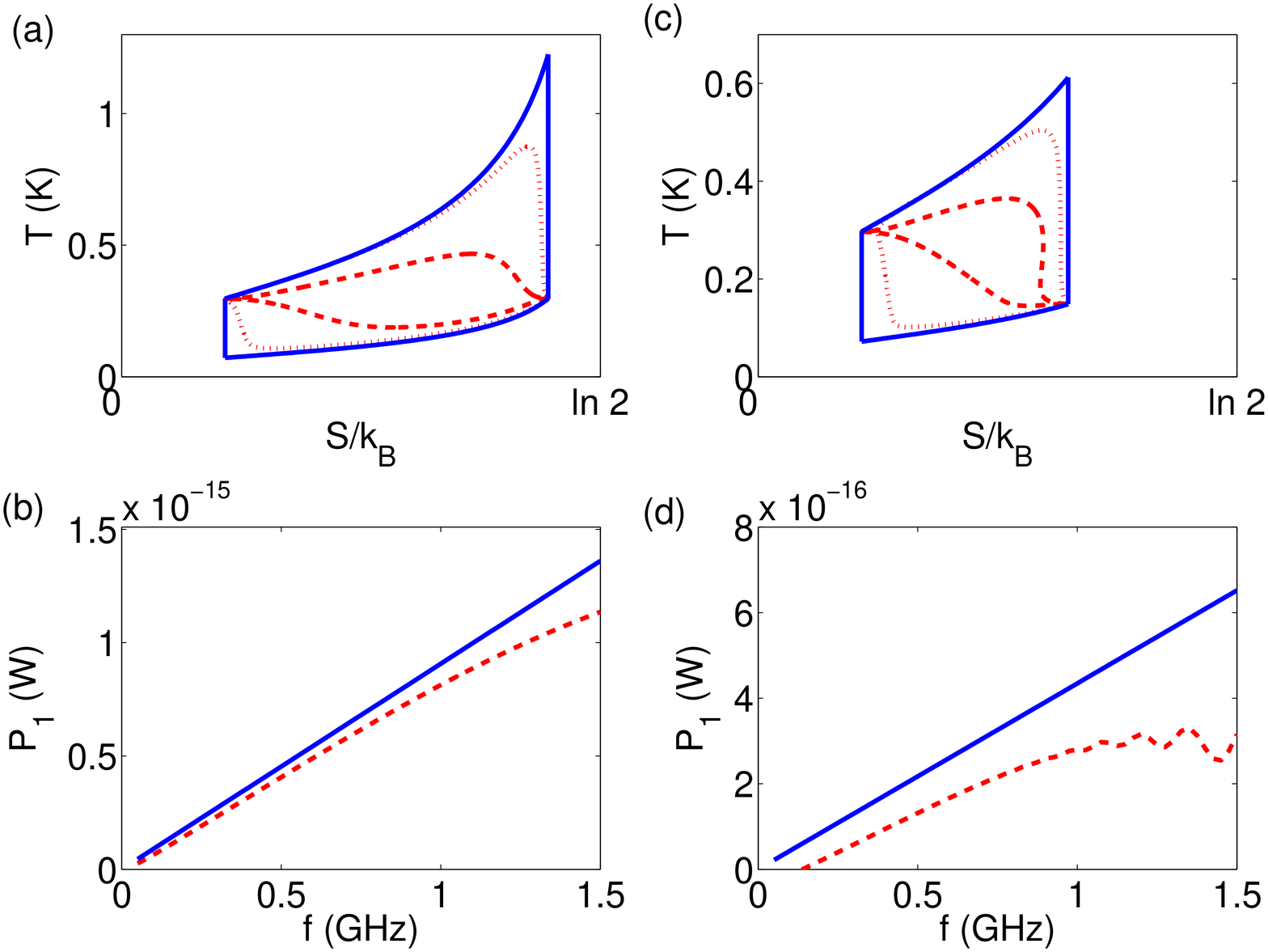}}
\end{picture}
\caption{\label{kuva2}
(color online)
Example of the simulated cooling power with $\omega_1/2\pi=\Delta/2\pi=5$ GHz ($\epsilon=0$ GHz), $\omega_2/2\pi=20.62$ GHz ($\epsilon=20$ GHz),
$Q_1=Q_2=10$, $\omega_j=\omega_{LCj}$
and $2(I_{\rm p}\Delta M_j)^2/(R_j\hbar\omega_j)=20\times 10^{9}s^{-1}$. This can be achieved e.g. with
$I_{\rm p}=200$ nA, $M_1=29$ pH, $M_2=59$ pH and  $R_1=R_2=1~\Omega$. The driving is sinusoidal.
(a) The solid line illustrates the path in the $T-S$ plane for the ideal cycle described in the text while the dashed (dotted)
line is a result of simulation for $f=0.05$ GHz ($f=1$ GHz) with $T_1=T_2=0.3\times\hbar\omega_2/k_{\rm B}\approx 300$ mK.
(b) Simulated cooling power vs. $f$ for the same temperatures as in (a) is shown with the dashed line while the solid line is the ideal result of
Eq.~\ref{eq:pow}
(c-d) Same as (a-b) but with $T_1=0.5\times T_2\approx 150$ mK. The cooling threshold at $0.14$ GHz in (d)
is caused by finite $Q$-factor. }
\end{figure}

To illustrate the practical potential of the device we show in
Fig.~\ref{kuva2} the simulated cooling power with sinusoidal driving
of $\epsilon(t)$ compared to the ideal case along with the actual
loop in the entropy temperature plane. The heat flow $P_j$ from
resistor $j$ to the qubit is simply obtained by integrating the
product of the thermalization rate and the energy deficit, i.e.,
$P_{j}=f\int_{0}^{1/f}dt\Gamma_{\rm th}^j \left[{\rm Tr} (\rho_{\rm
eq}(\beta_j,\epsilon(t)) H)-{\rm Tr} (\rho(t) H)\right]$ . The
density matrix $\rho(t)=\frac{1}{2}\vec{M}(t)\cdot\vec{\sigma}$ is
solved numerically using the Bloch equation (system is followed over
a few periods until it has converged to the limit cycle). We see
that the actual simulated behavior does not significantly deviate
at low $f$ from the ideal behavior and that cooling powers on the
order of fW can be achieved with reasonable sample parameters. The
oscillatory behavior at high $f$ is interpreted as Landau-Zener
interference \cite{sillanpaa,oliver}.

\begin{figure}
\begin{picture}(150,190)
\put(-60,0){\includegraphics[width=0.52\textwidth]{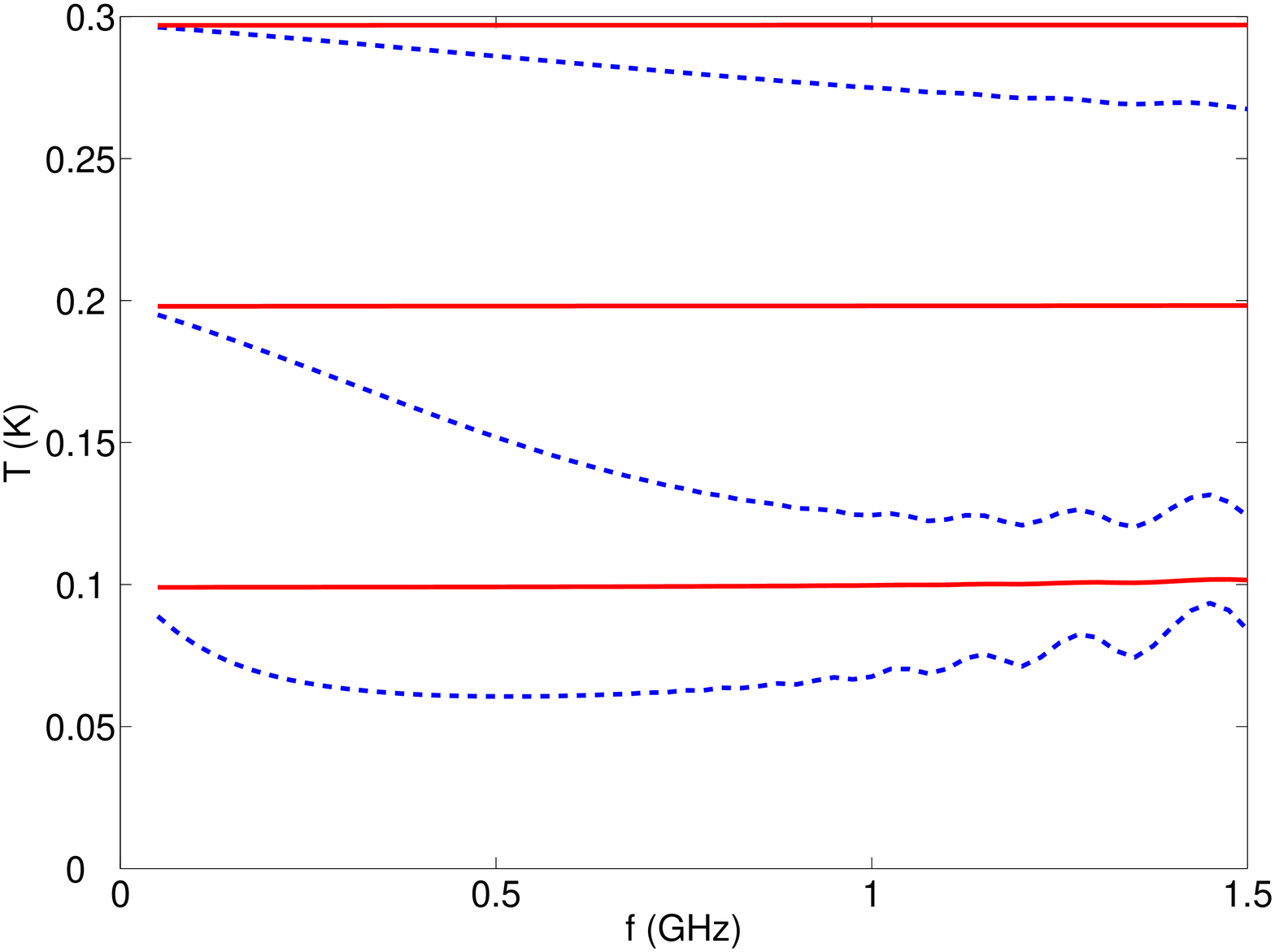}}
\end{picture}
\caption{\label{kuva3}
(color online) Equilibrium temperature as a function of pump frequency for three different phonon bath temperatures.
The temperature of resistor 1 (volume $10^{-21}$m$^3$) is shown with dashed line while the temperature of resistor 2
(volume $10^{-18}$m$^3$) is shown with
solid line. The bath temperatures $T_{\rm ph}\approx T_2$
from top to bottom are $0.3\times\hbar\omega_2/k_{\rm B}$, $0.2\times\hbar\omega_2/k_{\rm B}$  and $0.1\times\hbar\omega_2/k_{\rm B}$. Otherwise the parameters are
like in Fig.~\ref{kuva2}.}
\end{figure}

However, the cooling power has to be compared with realistic heat
loads to evaluate the utility of the flux qubit cooler. On one
hand, resistor 1 is subject to heat load from the phonons of the
substrate on which the device rests. On the other hand, resistor 2
should be coupled well enough to phonon bath such that the
unavoidable work done on it does not raise $T_2$ excessively. The
heat flow between the electron system of resistor $j$ and the phonon
system is given by $P_{el-ph}=\Sigma V(T^5_{j}-T_{ph}^5)$ where
$V_{j}$ is the volume of resistor $j$ and $\Sigma$ is typically on
the order of $10^{9}~{\rm Wm}^{-3}{\rm K}^{-5}$. Thus resistor 1
needs to have a sufficiently small volume while resistor 2 should be
large enough physically in order to serve as a heat sink. In
addition the photonic heat conduction between the resistors due to
temperature gradient may in principle contribute also. Following an
analysis similar to Ref.~\cite{schmidt}, the heat flow from
admittance $Y_2(\omega)$ to $Y_1(\omega)$ can be written as
\begin{equation}
\small
P_{\gamma}=\int_0^\infty\frac{d\omega}{2\pi}\left[4\hbar\omega^3M^2{\rm Re}Y_1(\omega){\rm Re}Y_2(\omega)
(n_2(\omega)-n_1(\omega))\right],
\end{equation}
where $n_j(\omega)=[\exp(\beta_j\hbar\omega-1)]^{-1}$ are the boson
occupation factors and $M$ is the mutual inductance between the
loops. For detuned high-Q resonators the photonic heat conduction
turns out to be quite negligible. For instance for the values of
Fig.~\ref{kuva3} with $M=5$ pH and $R_1=R_2=1~\Omega$ we get only
$P_{\gamma}=2\times 10^{-18}$ W even if $T_1=0$ K and $T_2=300$ mK.
Figure~\ref{kuva2} illustrates the calculated equilibrium
temperature versus operation frequency obtained numerically by
finding the balance between the dominating phononic heat conduction
and the integrated cooling power. We see that almost a factor of 2
reduction of $T_1$ is possible with realistic parameters.

In practice the drop of $T_1$ can be measured e.g. using an
additional SINIS thermometer, in which resistor 1 will serve as the
normal metal N. Its reading is sensitive to the electronic
temperature of N only, and self-heating can be made very small. The resistors should be
made out of thin film normal metal such as copper or gold with
typically sub 1 $\Omega$ square resistance. Volume can be picked
freely. To get the resonant frequencies and quality factor as above
we need $L_1=320$ pH, $C_1=3.2$ pF, $L_2=80$ pH and $C_2=0.8$ pF
which are also realistic. For the inductor one may use either
Josephson or the kinetic inductance of superconducting wire while the capacitance values are
similar to those in typical flux qubits \cite{yoshihara}. To
satisfy the conditions of the above numerical example we need quite
large mutual inductances which however can be easily achieved using
e.g. kinetic inductance \cite{prbspectro}. The strong driving
requires also rather large inductance between the microwave line and
the qubit, which should not result in uncontrolled relaxation. For
instance, $M_{\rm mw}$=5 pH coupling to the control line is
acceptable as it would result in at most $3\times 10^7 $ s$^{-1}$
relaxation rate assuming a 50 $\Omega$ environment at 0.3 K. This
choice will not degrade the performance of the device significantly
since driving is much faster. Yet sufficiently strong driving can be
achieved with a modest 3 $\mu$A ac current. 
Fabrication process will require most likely three lithography
steps.

In conclusion, we have described a method of  using a
superconducting flux qubit driven strongly at microwave frequency to
cool an external metal resistor. Here we considered LC resonators to
achieve the required frequency selectivity but a coplanar wave-guide
resonator or a mechanical oscillator could be used in principle,
too. We demonstrated by a numerical example that it is possible to
observe the associated temperature decrease experimentally. This
effect is directly related to the loss of information and thus to
the increase of entropy of the quantum bit.

\acknowledgments

J.P.P thanks NanoSciERA project "NanoFridge" of EU for
financial support.


\begin{thebibliography}{}
\bibitem{astafiev}
O. Astafiev, Yu. A. Pashkin, Y. Nakamura, T. Yamamoto, and J. S. Tsai,
Phys. Rev. Lett. {\bf 93}, 267007 (2004).
\bibitem{yoshihara}
F. Yoshihara, K. Harrabi, A. O. Niskanen, Y. Nakamura, J.S. Tsai, Phys. Rev. Lett. {\bf 97}, 167001 (2006).
\bibitem{bertet}
P. Bertet, I. Chiorescu, G. Burkard, K. Semba, C. J. P. M. Harmans, D.P. DiVincenzo, and J.E. Mooij,
Phys. Rev. Lett. {\bf 95}, 257002 (2005).
\bibitem{mooij}
J. E. Mooij, T. P. Orlando, L. Levitov, L. Tian,
C. H. van der Wal, S. Lloyd, Science {\bf 285}, 1036 (1999).
\bibitem{chiorescu}
I. Chiorescu, Y. Nakamura, C. J. P. M. Harmans, and J. E. Mooij,  Science {\bf 299}, 1869 (2003).
\bibitem{mit}
S. O. Valenzuela, W. D. Oliver, D. M. Berns, K. K. Berggren, L. S. Levitov, and T. P. Orlando,
Science {\bf 314}, 1589 (2006).
\bibitem{meschke}
M. Meschke, W. Guichard, and J. P. Pekola, Nature(London) {\bf 444}, 187 (2006).
\bibitem{hauss}
J. Hauss, A. Fedorov, C. Hutter, A. Shnirman, and G. Sch\"{o}n, cond-mat/0701041.
\bibitem{nisrfset}
J. P. Pekola, F. Giazotto, and O. P. Saira, Phys. Rev. Lett. {\bf 98}, 037201 (2007).
\bibitem{quan}
H.T. Quan, Y.X. Liu, C. P. Sun, and Franco Nori, quant-ph/0611275.
\bibitem{makhlin}
Yu. Makhlin, G. Sch\"{o}n, and A. Shnirman,
in {\it New Directions in Mesoscopic Physics (Towards Nanoscience)},
edited by R. Fazio, V. F. Gantmakher, and Y. Imry (Kluwer, 2003), p. 197; cond-mat/0309049.
\bibitem{sillanpaa}
M. Sillanp\"{a}\"{a}, T. Lehtinen, A. Paila, Yu. Makhlin, and P. Hakonen,
Phys. Rev. Lett. {\bf 96}, 187002 (2006).
\bibitem{oliver}
W. D. Oliver, Y. Yu, J. C. Lee, K. K. Berggren, L. S. Levitov, and T. P. Orlando,
Science {\bf 310}, 1653 (2005).
\bibitem{schmidt}
D. R. Schmidt, R. J. Schoelkopf, and A. N. Cleland,
Phys. Rev. Lett. {\bf 93}, 045901 (2004).
\bibitem{prbspectro}
A. O. Niskanen, K. Harrabi, F. Yoshihara, Y. Nakamura, and J. S. Tsai,
Phys. Rev. B {\bf 74}, 220503(R) (2006).

\end{thebibliography}
\end{document}